\documentclass[aps,prl,onecolumn,superscriptaddress,groupedaddress]{revtex4}   
\usepackage{graphicx}  
\usepackage{color, soul}
\usepackage{xcolor}
\usepackage{dcolumn}   
\usepackage{bm}        
\usepackage{amssymb}   
\usepackage{changepage}
\usepackage{amsmath}
\usepackage{hyperref}
\usepackage{float}
\usepackage{blindtext}
\hypersetup{
  colorlinks=false,
  linkbordercolor=blue
}

\usepackage[a4paper,bindingoffset=0.1in,%
            left=1.5cm,right=1.5cm,top=2cm,bottom=2.5cm,%
            footskip=0.5in]{geometry}
\usepackage{blindtext}

\begin{document}


\newlength{\halfpagewidth}
\setlength{\halfpagewidth}{\linewidth}
\divide\halfpagewidth by 2
\newcommand{\leftsep}{%
\noindent\raisebox{4mm}[0ex][0ex]{%
\makebox[\halfpagewidth]{\hrulefill}\hbox{\vrule height 3pt}}%
}
\newcommand{\rightsep}{%
\noindent\hspace*{\halfpagewidth}%
\rlap{\raisebox{-3pt}[0ex][0ex]{\hbox{\vrule height 3pt}}}%
\makebox[\halfpagewidth]{\hrulefill} } 
    
\title{The Origin of the Glass-like Thermal Conductivity in Crystalline Metal-Organic Frameworks}
\author{Yanguang~Zhou}
\email{maeygzhou@ust.hk}
\affiliation{Department of Mechanical and Aerospace Engineering, The Hong Kong University of Science and Technology, Clear Water Bay, Kowloon, Hong Kong}

\author{Yufei~Gao}
\affiliation{Key Laboratory of Ocean Energy Utilization and Energy Conservation of Ministry of Education, School of Energy and Power Engineering, Dalian University of Technology, Dalian, China}    

\author{Sebastian~Volz}
\affiliation{CNRS, UPR 288 Laboratoire d’Energétique Moléculaire et Macroscopique, Combustion (EM2C), Ecole Centrale Paris, Grande Voie des Vignes, 92295, Châtenay-Malabry, France}    
\affiliation{LIMMS/CNRS-IIS(UMI2820) Institute of Industrial Science, University of Tokyo 4-6-1 Komaba, Meguro-ku Tokyo, 153-8505, Japan}

\date{\today}

\begin{abstract}
It is textbookly regarded that phonons, i.e., an energy quantum of propagating lattice waves, are the main heat carriers in perfect crystals. As a result, in many crystals, e.g., bulk silicon, the temperature-dependent thermal conductivity shows the classical $1/T$ relationship because of the dominant Umklapp phonon-phonon scattering in the systems. However, the thermal conductivity of many crystalline metal-organic frameworks is very low and shows no, a weakly negative and even a weakly positive temperature dependence (glass-like thermal conductivity). It has been in debate whether the thermal transport can be still described by phonons in metal-organic frameworks. Here, by studying two typical systems, i.e., crystal zeolitic imidazolate framework-4 (cZIF-4) and crystal zeolitic imidazolate framework-62 (cZIF-62), we prove that the ultralow thermal conductivity in metal-organic frameworks is resulting from the strong phonon intrinsic structure scattering due to the large mass difference and the large cavity between Zn and N atoms. Our mean free path spectrum analysis shows that both propagating and non-propagating anharmonic vibrational modes exist in the systems, and contribute largely to the thermal conductivity. The corresponding weakly negative or positive temperature dependence of the thermal conductivity is stemming from the competition between the propagating and non-propagating anharmonic vibrational modes. Our study here provides a fundamental understanding of thermal transport in metal-organic frameworks and will guide the design of the thermal-related applications using metal-organic frameworks, e.g., inflammable gas storage, chemical catalysis, solar thermal conversion and so on.
\end{abstract}
\setulcolor{blue}
\maketitle
\twocolumngrid 

Metal-organic frameworks (MOFs), which are characterized by metal ions or clusters connected by organic bridges \cite{Li1999, Eddaoudi2002, Yaghi2003, Rowsell2004}, have attracted intensive attention in the applications of gas storage\cite{Rosi2003}, gas separation\cite{Rowsell2005}, and catalysis\cite{Lee2009} due to their extremely high porosity and internal surfaces areas. One important, and often neglected, challenge related to the gas storage in MOFs is the heat generation during the gas absorption process, which will cause the reduction of the gas absorption capacity\cite{Xiao2013, Babaei2018} and even accidents\cite{Woellner2018, Wang2018}. It has been suggested that the heat generated in the gas absorption process is mainly dissipated through the MOF frameworks\cite{Babaei2016}. Therefore, an in-depth understanding of thermal transport properties in MOFs is very important for optimizing the thermal design of these MOF-related structures and materials\cite{Xiao2013, Babaei2018, Babaei2016, Huang20071, Kim2017, Kalmutzki2018, Sezginel2018, Cui2018, Han2014, Zhang2013, Huang20072}. Unlike the thermal transport properties of many crystals, e.g., crystalline Si\cite{Glassbernner1964, Kremer2004}, crystalline Ge\cite{Glassbernner1964} and bulk diamond\cite{Slack1964}, which have been well-studied, whether the heat transfer mechanisms in these crystals are still suitable for the crystalline MOFs remains debatable and poorly documented. For example, both experiments\cite{Huang20071} and simulations\cite{Huang20072} show that the thermal conductivity of the metal-organic framework-5 (MOF-5) has a weak positive temperature dependence of $T^{0.13}$. The authors address this to the dual thermal transport channels including the contributions from both the anharmonic propagating and harmonic non-propagating vibrational modes which were depicted by the Cahill-Pohl model\cite{Cahill1992}. While Zhang et. al.\cite{Zhang2013} argue that the weak positive temperature dependence of the thermal conductivity in another MOF, i.e., ZIF-8, is stemming from enhanced overlap in the vibrational density of states between Zn and N atoms which is resulting from the overlap of trajectories between Zn and N atoms. Nevertheless, all these studies rely on the fact that the concept of phonons is still valid in MOFs. Despite these deductions in exploring the thermal transport mechanisms in MOFs, directly quantifying the scattering mechanisms of phonons and the modal level information, e.g., phonon mean free path and thermal conductivity contributed from various heat carriers in MOFs, is still lacking.

In this paper, the thermal transport properties are quantitatively analyzed in the example systems of crystalline zeolitic imidazolate framework (ZIF)-4 and ZIF-62, using atomistic simulations. For comparison, we also investigate the thermal transport mechanisms in crystalline silicon. Our analysis supports that, unlike anharmonic propagating modes are dominant heat carriers in crystals, both the propagating and non-propagating anharmonic vibrational modes are important heat carriers in crystalline MOFs and contribute largely to thermal energy transport. The non-propagating vibrational modes transfer heat energy via the overlap in the vibrational trajectory or equivalently to the vibrational density of states between Zn and N atoms and the resulting thermal conductivity has therefore a weak positive temperature dependence. The propagating vibrational modes contribute to the thermal conductivity through extremely strong phonon-phonon scattering, and the corresponding thermal conductivity thus holds a weak negative temperature dependence. The weak temperature dependence of the thermal conductivity in MOFs is a result of the two competitive mechanisms mentioned above. When the propagating vibrational modes are the main heat carriers in MOFs, the thermal conductivity shows a negative temperature dependence, e.g., ZIF-4 in our case. When the non-propagating vibrational modes are dominant, the thermal conductivity of the MOF will have a positive temperature dependence, e.g., ZIF-62 in our case. 

\begin{figure}
\hspace*{-5mm} 
\vspace*{-2mm} 
\setlength{\abovecaptionskip}{0.1in}
\setlength{\belowcaptionskip}{-0.2in}
\includegraphics [width=3.2in]{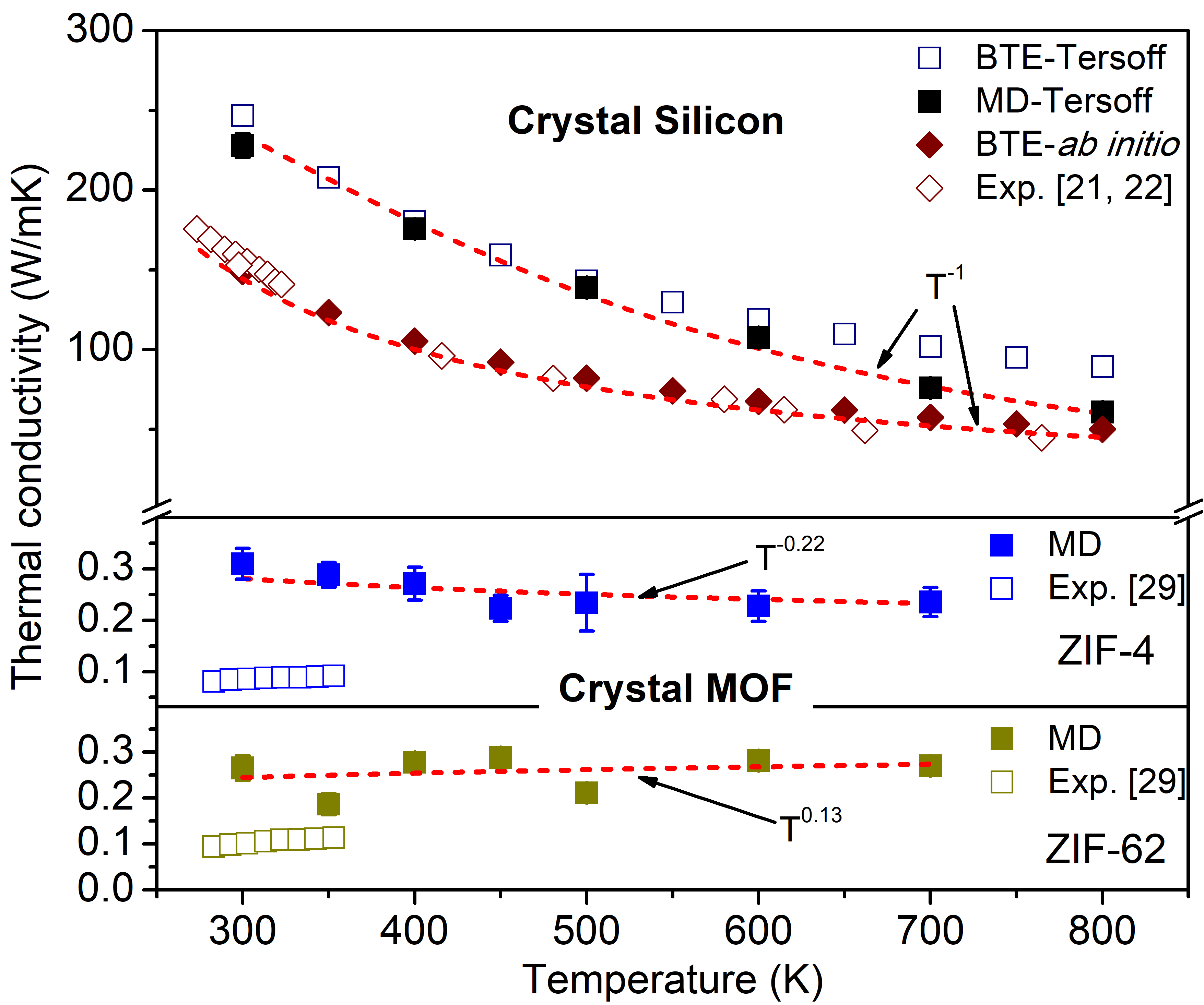}
\caption{The temperature-dependent thermal conductivity of crystalline silicon (upper panel), crystal ZIF-4 (middle panel) and ZIF-62 (lower panel). The experimental data of crystalline silicon is addressed from\cite{Glassbernner1964, Kremer2004}, and the measurements of crystal ZIF-4 and ZIF-62 are taken from reference\cite{Sorensen2020}.}
\label{fig:F1}
\end{figure}

In the crystalline ZIF-4 and ZIF-62, referred as to cZIF-4 and cZIF-62 thereafter, each Zn$^{2+}$ atom forms a tetrahedron by linking to four N atoms of the organic groups. The lattice constants of the unit cell for cZIF-4 and cZIF-62 are 1.52 nm and 1.65 nm, respectively. The density of cZIF-4 and cZIF-62 is 1.20 g/cm$^3$ and 1.18 g/cm$^3$, respectively. The density of cZIF-4 in our case is slightly lower than the crystallographic density 1.22 g/cm$^3$\cite{Gaillac2017}, and other reported values of 1.28 g/cm$^3$\cite{Yang2018} and 1.25 g/cm$^3$\cite{Gaillac2017}. The difference between our values and other existing results may exist because: i) our simulations are implemented in an environment with standard atmospheric pressure and ii) the effect of temperature and potential are addressed in our systems. For comparison, we also calculate the thermal transport properties of crystalline silicon, indicated as cSi thereafter, using both first-principles (FP) and classical molecular dynamics (MD) simulations. The lattice constants in FP and MD are 0.538 nm and 0.544 nm, respectively. 

\textbf{Figure ~\ref{fig:F1}} shows the temperature-dependent thermal conductivity of cSi, cZIF-4, and cZIF-62. The thermal conductivities of cSi calculated using both FP and MD show a typical $1/T$ relationship which stems from the dominant Umklapp ($U$) phonon processes\cite{Broido2005}. We also find that both the classical potential and the FP thermal conductivities computed using Boltzmann transport equation (BTE) are a little lower than the values of MD calculations and experimental measurements\cite{Glassbernner1964, Kremer2004} at high temperatures, e.g., above 600 K, respectively. The reason for this is that we only include the three-phonon scattering processes in the BTE calculations, in which the high-order phonon scattering processes included in MD and experiments inherently should be also considered at high temperatures\cite{Feng2016}. 

We next move to the temperature-dependent thermal conductivity of cZIF-4 and cZIF-62. Due to the super large size of the unit cells of MOFs, the MD simulation may be the only accessible approach to consider their thermal transport properties here. The thermal conductivities of both cZIF-4 and cZIF-62 show weak temperature dependences, which agree with experimental observations well. However, our MD results of cZIF-4 and cZIF-62 are much higher than measured values of thermal conductivities\cite{Sorensen2020}, and our results of cZIF-4 even show a different temperature dependence compared to the experimental measurements. These differences can be addressed from several aspects: i) the structures of the MOFs in MD simulations are fully symmetric, while the experimental samples possess some moiety sites of partial occupancy which can scatter the lattice vibrations; ii) there are defects and air entrapment in the samples while MD simulates the ideal structures; iii) the errors stemming from the atomic potential and the process of the measurements. The contributions to the thermal conductivity from various heat carriers are changed by these factors, which then lead to the different temperature dependence (see discussions below for details). We also emphasize that the inaccuracy resulting from the classical distribution in MD simulations (Boltzmann vs. Bose-Einstein) can be ignored since the Debye temperatures of cZIF-4 and cZIF-62 are much lower than the simulated temperatures (see supporting materials\cite{SI2020}). 

\begin{figure*}
\hspace*{-4mm} 
\vspace*{-2mm} 
\setlength{\abovecaptionskip}{0.2in}
\setlength{\belowcaptionskip}{-0.2in}
\includegraphics [width=5.5in]{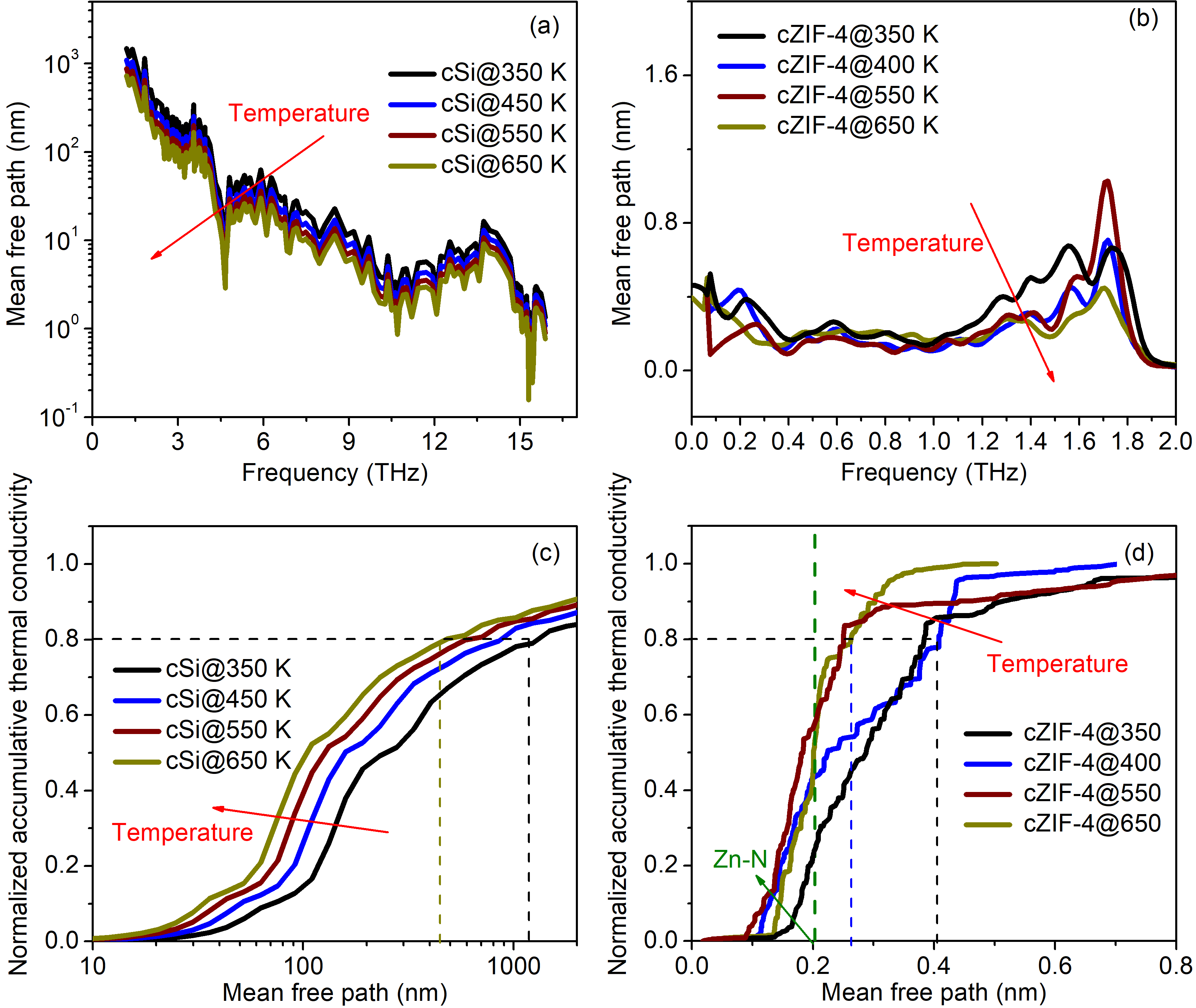}
\caption{The mean free path spectrum of (a) crystal silicon and (b) cZIF-4 at different temperatures, and the accumulative thermal conductivity of (c) crystalline silicon and (d) cZIF-4 at the corresponding temperatures mentioned above. The dashed lines are eye guidance.}
\label{fig:F2}
\end{figure*}

Mean free path analysis: To gain insight in the physical mechanisms leading to the abnormal temperature dependence of the thermal conductivity observed in cZIF-4 and cZIF-62, we calculate the thermal conductivity spectrum and the corresponding mean free path (MFP) of the example systems, i.e., cZIF-4 at various temperatures, using the spectral thermal conductivity analysis. For comparison, we also calculate the corresponding values of cSi at different temperatures. 

Our spectrum results (\textbf{Figure ~\ref{fig:F2}a}) show that the mean free path (MFP) of cSi can be as large as several micrometers at 350 K, and decreases quickly with regard to temperature due to the stronger U scatterings, e.g., the maximal MFP decreases from 2 um at 350 K to 800 nm at 650 K. For the cZIF-4 (\textbf{Figure ~\ref{fig:F2}b}), although the MFP spectrum distribution is not largely changed by the temperature, the MFPs are found to be quite small and these lattice vibrations with frequencies above 1 THz are decreasing with the temperature somehow. It is known that the internal cavities occupy a quite large volume in MOFs. As a result, the lattice vibrations are strongly scattered by these internal cavities (see detailed discussion below), which lead to the short MFPs (\textbf{Figure ~\ref{fig:F2}d}) and the ultralow thermal conductivity (\textbf{Figure ~\ref{fig:F1}}) as observed here. It is also not surprising to find that these vibrations with frequency smaller than about 2 THz are the main heat carriers in cZIF-4. In solids, high-frequency vibrations normally contribute much less to the heat exchange compared to the low-frequency vibrations due to their larger scattering rates and flatter dispersions.

\begin{figure}
\hspace*{-5mm} 
\vspace*{-2mm} 
\setlength{\abovecaptionskip}{0.1in}
\setlength{\belowcaptionskip}{-0.2in}
\includegraphics [width=3.5in]{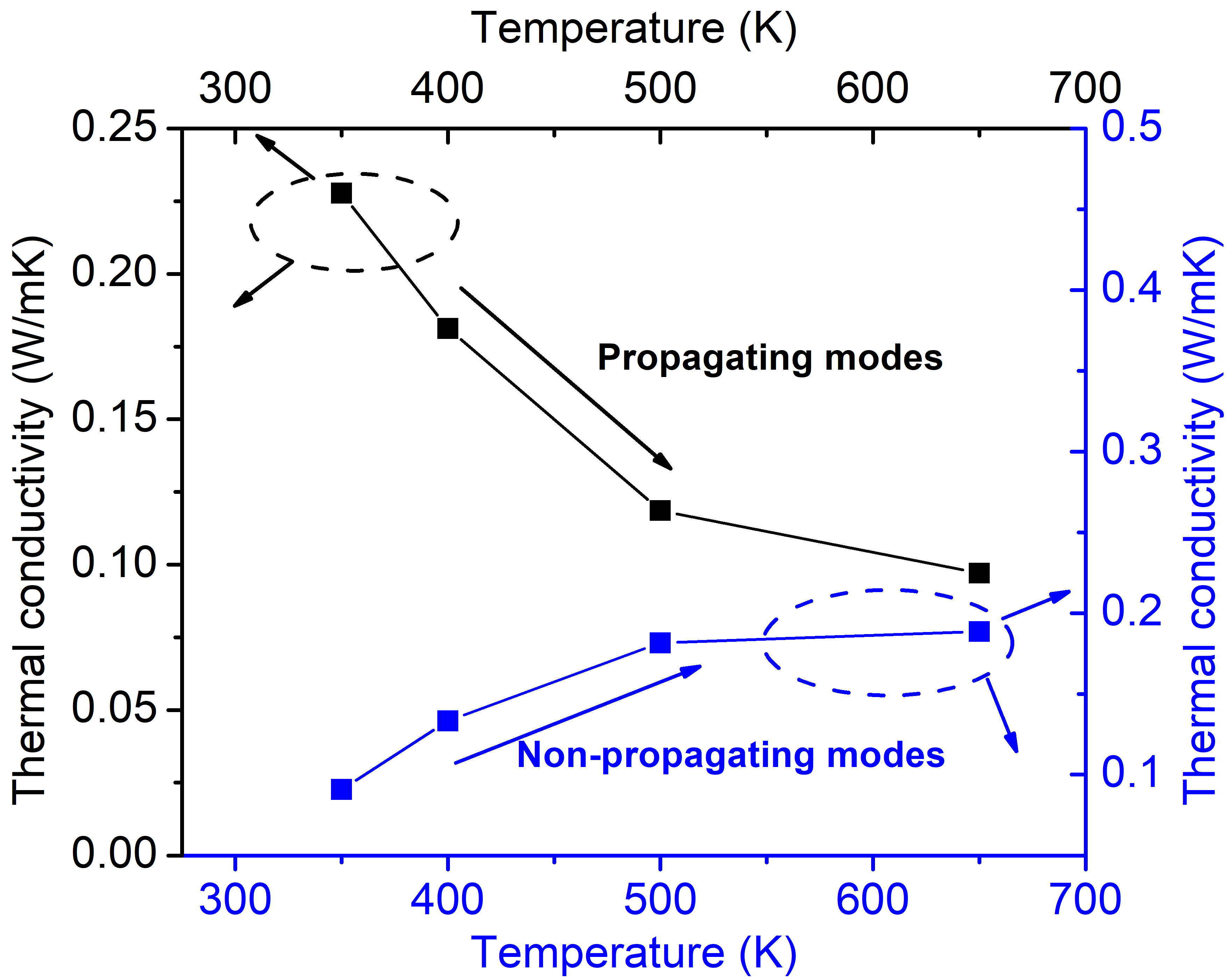}
\caption{The temperature-dependent thermal conductivity of propagating vibrational modes and non-propagating vibrational modes.}
\label{fig:F3}
\end{figure}

\begin{figure*}
\hspace*{-4mm} 
\vspace*{-2mm} 
\setlength{\abovecaptionskip}{0.1in}
\setlength{\belowcaptionskip}{-0.2in}
\includegraphics [width=5.5in]{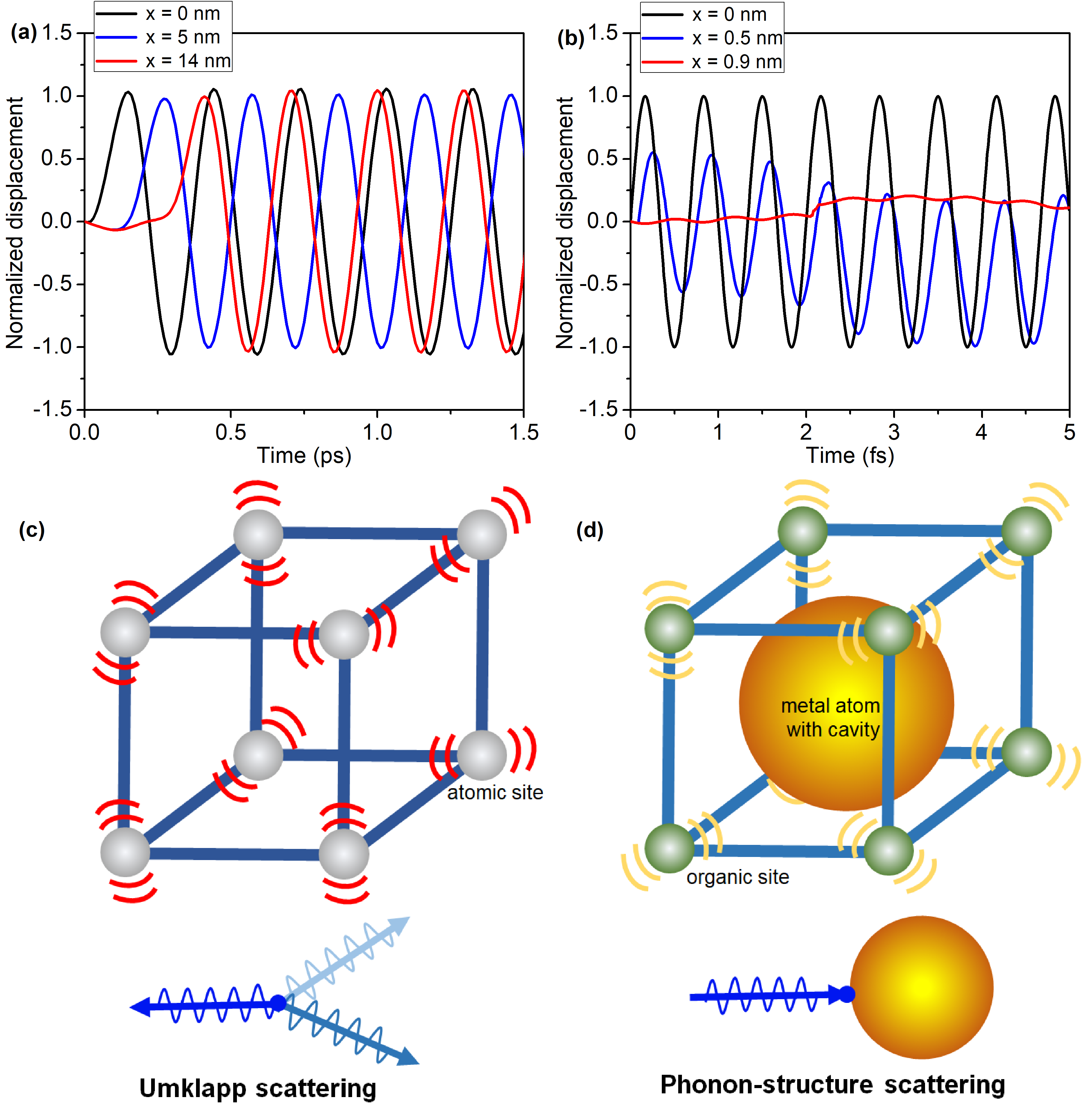}
\caption{The wave decay process in (a) crystal silicon and (b) crystal ZIF-4, and the corresponding scattering picture in (c) crystal silicon and (d) crystal ZIF-4.}
\label{fig:F4}
\end{figure*}

To quantitatively characterize the contribution of the heat carriers in the systems, the thermal conductivity accumulation function $\kappa (\Lambda )$ of vibrational modes has been computed from the MFP $\Lambda $ using
\vspace*{-2mm} 
\begin{equation}
\vspace*{-2mm} 
\kappa ({{\Lambda }_{0}})=\sum\limits_{\Lambda <{{\Lambda }_{0}}}{\kappa (\Lambda )}
\label{eqn:A1}  
\end{equation}

Our results (\textbf{Figure ~\ref{fig:F2}c}) show that the MFP of heat carriers in cSi is quite broad, i.e., from 10 nm to 1 um, and fastly decreases with temperature due to the enhanced scattering among heat carriers. While for cZIF-4 (\textbf{Figure ~\ref{fig:F2}d}), the MFP is quite small and ranging from 0.1 nm to 0.6 nm because of the strong scattering inherent to cavity vibrations (see detailed analysis below). At the same time, our results (Figure 2d) show that the vibration modes with MFP smaller than 0.1 nm contribute little to the thermal energy exchange in cZIF-4. The vibration modes with MFP ranging from 0.3 nm to 0.4 nm which contribute to around 40$\%$ of the total thermal conductivity for cZIF-4 at 350 K, are scattered strongly when the system temperature is increased from 350 K to 650 K. Furthermore, the vibrational modes have MFPs smaller than 0.21 nm (green dashed line in \textbf{Figure ~\ref{fig:F2}d}), i.e., the length of Zn-N bond, and are therefore regarded as diffuson-like vibrational modes, contributing bby a large amount to the thermal transport and their contribution to thermal conductivity is also found to increase with temperature.   

\begin{figure*}
\hspace*{-4mm} 
\setlength{\abovecaptionskip}{0.1in}
\setlength{\belowcaptionskip}{-0.1in}
\includegraphics [width=5.5in]{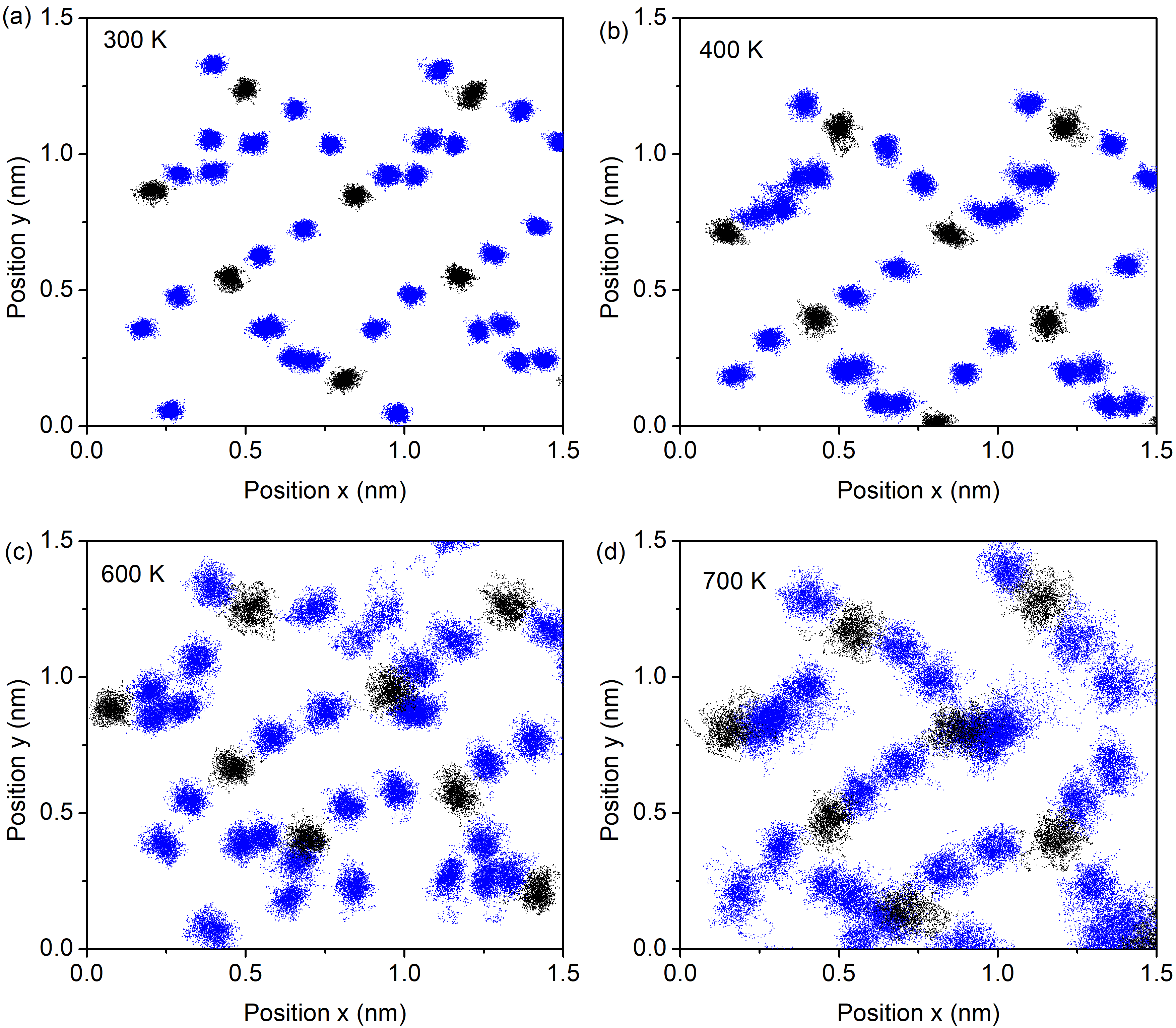}
\caption{The trajectories of Zn (black) and N (blue) atoms of crystal ZIF-4 at (a) 300 K, (b) 400 K, (c) 600 K and (d) 700 K.}
\label{fig:F5}
\end{figure*}

Quantifying the contribution to thermal conductivity from various vibrational modes: We now characterize the thermal energy carried by different heat carriers. In cSi, all the heat carriers have MFP larger than the minimal atomic distance of 0.24 nm, and therefore only propagating vibrational modes, i.e., phonons, are existing in the systems (\textbf{Figure ~\ref{fig:F2}c}). For cZIF-4, as discussed above, both propagating and non-propagating vibrational modes, i.e., propagons and diffusons, are existing and contributing largely to the thermal conductivity (\textbf{Figure ~\ref{fig:F2}d} and \textbf{Figure ~\ref{fig:F3}}). It is not surprising to find that the thermal conductivity contributed by the propagating vibrational modes decreases with temperature since the increase in temperature enhances the strength of the $U$ scattering process in the system. The thermal conductivity contributed by the non-propagating vibrational modes, i.e., diffusons, however, are found to increase with temperature. Based on the definition of diffusons proposed by Allen and Feldman [31], the non-propagating vibrational modes transfer thermal energy through the harmonic coupling between them, and therefore, the resulting thermal conductivity should be temperature independent when all the vibrational modes are active which should be the case in molecular dynamics simulations. Here, we emphasize that the diffusons in our systems are stemming from the overlap of the vibration\lq s trajectories between Zn and N atoms, which are different from the diffusons defined in the disordered solid systems by Allen and Feldman\cite{Allen1993}. Therefore, the thermal conductivity contributed from the diffusons in cZIF-4 shows a strong temperature dependence since the overlap of vibrations\lq trajectories or vibrational density of states (see the analysis for details below) is increasing with temperature. Such a phenomenon has also been widely observed in the partial-liquid partial-solid systems\cite{Zhou2018, Zhou2020}. The temperature dependence of the global thermal conductivity in MOFs is a result of the competition between the propagating and non-propagating vibrational modes: the thermal conductivity will decrease with temperature if propagating modes are the main heat carriers in the system and reversely if non-propagating modes predominate.       

To further answer the reason on the ultralow and abnormal thermal conductivity of MOFs, we record the decay process of a specific vibrational mode in both cSi and cZIF-4, i.e., $\omega =3.395\ \operatorname{THz}$ for cSi and $\omega =0.5\ \operatorname{THz}$ for cZIF-4, using the trigger wave method. In the trigger wave simulations, the anharmonicity can be ignored since the system temperature is set as 0.1 K. Our results show that the vibration in cSi is the one of a harmonic oscillator (\textbf{Figure ~\ref{fig:F4}a}), while the vibrations in cZIF-4 are strongly scattered by the intrinsic structure and transfer energy only over a few angstroms (\textbf{Figure ~\ref{fig:F4}b}) which agrees quite well with our MFP calculations above. Based on our analysis, only the anharmonic phonon-phonon scatterings are existing in cSi (\textbf{Figure ~\ref{fig:F4}c}), and thus the corresponding vibrations do not decay with time in low-temperature systems where the anharmonicity can be ignored. However, for the MOFs, e.g., cZIF-4 in our case, the vibrations are scattered strongly when they transfer thermal energy from the organic sites to the metal atoms because of the large mass difference between the metal atoms and the organic atoms, i.e., N atoms for ZIFs, and because of the large cavity between the metal atoms and organic sites (\textbf{Figure ~\ref{fig:F4}d}). Consequently, the thermal conductivity of MOFs is generally quite low, i.e., below 1 W/mK, as observed in our study and many others\cite{Babaei2016, Huang20071, Han2014, Zhang2013, Huang20072, Sorensen2020, Babaei2020}.

Next, we plot the vibrations\lq trajectories of Zn and N atoms in \textbf{Figure ~\ref{fig:F5}} at four typical temperatures, i.e., 300 K, 400 K, 600 K and 700 K. Our results show that the trajectories’ overlap between Zn and N atoms rapidly increases with temperature. Therefore, the thermal conductivity contributed by the heat carriers with MFPs shorter than the Zn-N distance, i.e., diffusons, is increasing with temperature as observed in \textbf{Figure ~\ref{fig:F3}}. When those diffusons are the dominant heat carriers in MOFs, the total thermal conductivity of MOFs increases with temperature which is shown in our results, i.e., cZIF-62, and has been widely observed in experimental investigations\cite{Huang20071, Sorensen2020} and other simulations [Zhang2013, Huang20072]. On the other side, if these diffusons are not the main heat carriers in the system, the total thermal conductivity of MOFs will follow the behavior of the thermal conductivity contributed by the propagating vibrational modes, i.e., the negative temperature dependence, which is observed in cZIF-4 in our paper and is found in experimental \cite{Erickson2015, Gunatilleke2017} and simulation\cite{Wang2015} works targeting MOFs.   

In conclusion, by performing ReaxFF atomistic simulations and spectral heat current analysis, we show that the ultralow thermal conductivity in crystalline metal-organic frameworks is resulting from the strong vibration-inherent structure scatterings due to the large mass difference and the large cavity between the metal atoms and the organic sites. The thermal conductivity contributed by the anharmonic propagating vibrational modes shows a negative temperature dependence due to the stronger scatterings among the vibrations in the higher temperature systems. However, the thermal conductivity resulted from the anharmonic non-propagating vibrational modes shows a positive temperature dependence because of the increasing trajectories’ overlap between the metal atoms and the organic sites. As a result, the temperature dependence of thermal conductivity for the crystalline metal-organic frameworks depends on the competition between the anharmonic propagating vibrations and the anharmonic non-propagating vibrational modes. When the propagating vibrational modes are the dominant heat carriers in the metal-organic frameworks, the thermal conductivity shows a weak negative temperature dependence. If the non-propagating anharmonic vibrational modes are the main heat carriers in the metal-organic frameworks, the total thermal conductivity follows a positive temperature dependence. Our results here support a quantitative picture for the abnormal temperature dependence of the thermal conductivity observed in metal-organic frameworks, which should guide the design of the future thermal-related applications.   

Y.Z. thanks startup fund (REC20EGR14 and a/c-R9246) from Hong Kong University of Science and Technology (HKUST).

\end{document}